\documentclass[12pt]{article}
\pdfoutput=1
\usepackage{amsbsy}
\usepackage{amsfonts}
\usepackage{amsmath,amssymb}
\usepackage{bbold}
\usepackage{pdfpages}

\newcommand{\sect}[1]{\setcounter{equation}{0}\section{#1}}

\newcommand{\eq}{\begin{equation}}
\newcommand{\eqa}{\begin{eqnarray}}  
\newcommand{\en}{\end{equation}}
\newcommand{\ena}{\end{eqnarray}}
\newcommand{\enn}{\nonumber \end{equation}}

\def\sk{\vskip .4cm}
\def\noi{\noindent}
\def\om{\omega}

\def\ga{\gamma}

\let \part\partial

\def\unquarto{{1 \over 4}}

\def\unmezzo{{1 \over 2}}
\def\epsi{\varepsilon}

\def\de{\delta}

\def\part{\partial}

\def\sk{\vskip .4cm}

\def\noi{\noindent}

\def\X0{X^0}

\def\om{\omega}

\def\ga{\gamma}

\def\unquarto{{1 \over 4}}
\def\unmezzo{{1 \over 2}}
\def\epsi{\varepsilon}

\def\de{\delta}

\def\Dcal{{\cal D}}

\def\Mcal{{\cal M}}

\def\square{{\,\lower0.9pt\vbox{\hrule \hbox{\vrule height 0.2 cm
\hskip 0.2 cm \vrule height 0.2 cm}\hrule}\,}}

\def\epsilonbar{{\bar \epsilon}}
\def\thetabar{{\bar \theta}}

\def\psibar{\bar \psi}
\def\epsilonbar{\bar \epsilon}
\def\chibar{\bar \chi}
\def\rhobar{\bar \rho}

\def\Sigmabar{\overline \Sigma}

\def\onebold{{\bf 1}}

\def\Gbb{\mathbb{G}}

\def\dright{\stackrel{\rightarrow}{\part}}
\def\dleft{\stackrel{\leftarrow}{\part}}

\def\Phibar{{\overline \Phi}}
\def\xibar{{\overline \xi}}

\begin{document}

\begin{titlepage}
\rightline{ARC-2020-08}
\vskip 2em
\begin{center}
{\Large \bf Covariant hamiltonian for supergravity \\ in $d=3$ and $d=4$} \\[3em]

\vskip 0.5cm

{\bf
Leonardo Castellani}
\medskip

\vskip 0.5cm

{\sl Dipartimento di Scienze e Innovazione Tecnologica
\\Universit\`a del Piemonte Orientale, viale T. Michel 11, 15121 Alessandria, Italy\\ [.5em] INFN, Sezione di 
Torino, via P. Giuria 1, 10125 Torino, Italy\\ [.5em]
 Arnold-Regge Center, via P. Giuria 1, 10125 Torino, Italy
}\\ [4em]
\end{center}

\begin{abstract}
\sk

We extend the covariant canonical formalism recently discussed in ref. [1] to geometric theories coupled to both bosonic and fermionic $p$-forms. This allows a covariant hamiltonian treatment of supergravity theories. As examples we present the
covariant hamiltonian formulation for $d=3$ anti-De Sitter supergravity and for the ``new minimal" $d=4$ , $N=1$ supergravity 
(with $1$-form and $2$-form auxiliary fields).  Form-Poisson brackets and form-Dirac brackets are defined, and used to find the form-canonical generators of all gauge symmetries via an algorithmic procedure.

\end{abstract}

\vskip 7cm
 \noi \hrule \vskip .2cm \noi {\small
leonardo.castellani@uniupo.it}

\end{titlepage}

\newpage
\setcounter{page}{1}

\tableofcontents

\vfill\eject

\sect{Introduction}

We have re-examined in ref. \cite{CD} the covariant hamiltonian approach proposed long ago by D' Adda, Nelson and Regge \cite{CCF1}-\cite{CCF5}, and extended it with a new definition of form-Poisson brackets and form-Dirac brackets consistent with the 
form-Legendre transformation that defines momenta and Hamiltonian. In short, the momenta $\pi^i$ are defined as
the derivative of a $d$-form Lagrangian $L$ with respect to the exterior derivative $d \phi_i$ of the fundamental $p$-form fields, and the $d$-form Hamiltonian is $H=(d\phi_i) \pi^i - L$. 

Here we generalize the covariant hamiltonian framework to include fermions, and apply it to $d=3$ and $d=4$ supergravity.

This framework is well suited to translate in hamiltonian language the group-geometric approach to supergravity developed since the late 70's, based on free differential algebras (FDA's) \cite{gm11}-\cite{gm23} (for a recent review see for ex. \cite{LC2018}). A comprehensive
account on Regge works can be found in the book \cite{reggebook}.

Hamiltonian methods are very useful for conceptual and practical reasons, prominent ones being study of the symmetries and quantization.
Our treatment being covariant (no time direction singled out), it simplifies considerably the canonical analysis and provides
a basis for a covariant quantization procedure. 

Other covariant Hamiltonian formalisms have been proposed in the literature, and 
a very partial list of references on multimomentum and multisymplectic canonical frameworks is given in ref. \cite{CD}, to which
we add the recent paper of \cite{Nakajima}. The essential ideas appeared in papers by De Donder and Weyl more than seventy years ago  \cite{DeDonder,Weyl}. Some of these approaches are quite similar in spirit to the one we discuss here, but to our knowledge the first proposal of a $d$-form Hamiltonian, together with its application to gravity, can be found in ref. \cite{CCF1}. 

In the present paper a definition of form Poisson brackets and form Dirac brackets for bosonic and fermionic fundamental $p$-forms is given, and applied to $d=3$ and $d=4$ supergravity to find all the canonical symmetry generators. The method allows to discover in a 
systematic way all the gauge symmetries of a theory reformulated in hamiltonian terms. Geometric theories have also 
an {\it a priori} symmetry by construction, i.e. diffeomorphism invariance. In $d=3$ pure gravity 
diffeomorphisms coincide with gauge translations, when we consider the theory in second order formalism with vanishing torsion. 
This is no more true for $d=3$ supergravity, and in higher dimensions. In these cases the transformation rules under diffeomorphisms 
can be obtained as usual by acting on the fields (and their momenta) with the Lie derivative.

In Section 2 we present a short r\'esum\'e of the covariant hamiltoniam formalism, based on ref. \cite{CD}, and apply it in Section 3 to $d=3$, $N=1$ AdS supergravity
and in Section 4 to $d=4$, $N=1$ supergravity with the auxiliary fields of the ``new minimal" model. One of the auxiliary fields being
a $2$-form, this example shows the versatility of the form-Poisson (and Dirac) brackets to accomodate $p$-forms in their definition.
Section 5 contains some conclusions, and $\gamma$ matrix conventions are summarized in the Appendices.

\sect{A summary of the covariant hamiltonian formalism} 

\subsection{Geometric action and Euler-Lagrange equations}

Consider the action :
\eq
S= \int_{\Mcal^d} L (\phi_i, d\phi_i)
\en
where the Lagrangian $L$ depends on a collection of $p_i$-form fields $\phi_i$ and their exterior derivatives, and is
integrated on a $d$-dimensional manifold $\Mcal^d$.

The variational principle reads
\eq
\delta S =  \int_{\Mcal^d} \delta \phi_i { \dright L \over \partial \phi_i} + d (\delta \phi_i ) { \dright L \over \partial (d\phi_i)}=0
\en
All products are exterior products between forms, satisfying
\eq
A B = (-)^{ab + \eta_a \eta_b} BA
\en
with $a,b$ and $\eta_a,\eta_b$ the form and fermionic gradings of the forms $A$ and $B$ respectively
($\eta = 0$ for bosons and $\eta = 1$  for fermions). The symbol ${ \dright L \over \partial \phi_i}$ indicates the right derivative of $L$ 
 with respect to a $p$-form $\phi_i$, defined by first bringing $\phi_i$ to the left in $L$
(taking into account the sign changes due to the gradings)
and then canceling it against the derivative. 

The corresponding Euler-Lagrange equations are:
\eq
d ~ { \dright L \over \partial (d\phi_i)} - (-)^{p_i} { \dright L \over \partial \phi_i} =0 \label{ELeqs}
\en

\subsection{Form Hamiltonian}

We define the $d$-form Hamiltonian density as:
\eq
H \equiv d\phi_i ~\pi^i  - L \label{formH}
\en
where the ($d-p_i-1$)-form momenta are given by:
\eq
\pi^i \equiv {\dright L \over \partial (d\phi_i)} \label{momentadef}
\en
The form-analogue of the Hamilton equations reads:
\eq
d \phi_i = (-)^{(d+1)(p_i+1)+ \eta_i} ~ {\dright H \over \partial \pi^i} ,~~~d \pi^i =  (-)^{p_i+1} ~{\dright H \over \partial \phi_i} \label{formHE}
\en
The first equation is equivalent to the momentum definition, and is obtained by taking the right derivative of $H$ as given in 
(\ref{formH}) with respect to $\pi^i$:
\eq
{\dright H \over \partial \pi^i} = {\dright d\phi_j \over \partial \pi^i} ~\pi^j + (-)^{(d-p_i - 1)(p_i+1)+ \eta_i} ~ d\phi_i - 
 {\dright d\phi_j \over \partial \pi^i} ~ {\dright L \over \partial (d\phi_j)}
\en
and then using (\ref{momentadef}), and $(d-p_i-1)(p_i+1) = (d+1)(p_i + 1) (mod~ 2)$.

The second is equivalent to the Euler-Lagrange form equations since 
\eq
{\dright H \over \partial \phi_i} = {\dright d\phi_j \over \partial \phi_i} ~\pi^j - { \dright L \over \partial \phi_i} -  {\dright d\phi_j \over \partial \phi_i}~ {\dright L \over \partial (d\phi_j)} = - { \dright L \over \partial \phi_i} 
\en
because of the momenta definitions (\ref{momentadef}). Then using (\ref{ELeqs}) yields the form Hamilton equation for $d\pi^i$.

\subsection{Form Poisson bracket}

The form Hamilton equations allow to express the (on shell) exterior differential of any $p$-form $F(\phi_i, \pi^i)$ as
\eq
dF=d\phi_i ~{\dright F\over \partial \phi_i} + d \pi^i ~{\dright F\over \partial \pi^i} =  (-)^{(d+1)(p_i+1)+\eta_i} ~{\dright H \over \partial \pi^i}~
{\dright F\over \partial \phi_i} +  (-)^{p_i+1} ~{\dright H \over \partial \phi_i} {\dright F\over \partial \pi^i} 
\en
Using left derivatives this expression simplifies:
\eq
dF= {\dleft H \over \partial \pi^i}~
{\dright F\over \partial \phi_i} -  (-)^{p_i d+ \eta_i} ~{\dleft H \over \partial \phi_i} {\dright F\over \partial \pi^i}  \label{differential}
\en
{\bf Note:} left derivatives are defined as ``acting on the left" and for example ${\dleft H \over \partial \phi_i}$ really means
${H \dleft \over \partial \phi_i}$. It is easy to verify\footnote{suppose that $A$ is contained in $F$ as $F= F_1 A F_2$. Then
${\dright F\over \partial A}  = (-)^{af_1+ \eta_a \eta_{f_1}} F_1 F_2$ and ${\dleft F\over \partial A}  = (-)^{af_2+ \eta_a \eta_{f_2}} F_1 F_2$ so that ${\dleft F\over \partial A} =
(-)^{a(f_1+f_2)+ \eta_a (\eta_{f_1}+ \eta_{f_2} )} {\dright F\over \partial A} = (-)^{a(f-a)+ \eta_a (\eta_f -\eta_a)} {\dright F\over \partial A} $ and (\ref{leftright}) follows.}
that the left and right derivatives of an $f$-form $F$ with respect 
to an $a$-form $A$ satisfy
\eq
{\dleft F \over \partial A} = (-)^{a(f+1)+ \eta_a (\eta_f +1)} ~{\dright F \over \partial A} \label{leftright}
\en
and this relation is used to prove eq. (\ref{differential}).
\sk

The expression for the differential (\ref{differential}) suggests the definition of the {\it form Poisson bracket} (FPB):
\eq
\{ A, B \} \equiv  {\dleft B \over \partial \pi^i}~
{\dright A\over \partial \phi_i} -  (-)^{p_i d+ \eta_i} ~{\dleft B \over \partial \phi_i} {\dright A\over \partial \pi^i}  \label{FPB}
\en
so that
\eq
dF = \{ F,H \} \label{differential2}
\en
The form Poisson bracket between the $a$-form $A$ and the $b$-form $B$ is a ($a+b-d+1$)-form, and canonically conjugated forms satisy:
\eq
 \{ \phi_i, \pi^j\}  = \delta_i^j \label{canonicalPB}
 \en

As observed in ref. \cite{CCF5}, there is a convenient notation that encodes both form and fermionic gradings.
Defining the vectors
\eq
[\phi_i] \equiv (p_i, \eta_i),~~[\pi^i] \equiv (d-p_i-1, \eta_i)
\en
where the first components are the form gradings of $\phi_i$ and $\pi^i$, and the second components are their
fermionic gradings, the form Poisson bracket can be rewritten as:
\eq
\{ A, B \} \equiv  {\dleft B \over \partial \pi^i}~
{\dright A\over \partial \phi_i} -  (-)^{[\phi_i] \cdot[\pi^i]} ~{\dleft B \over \partial \phi_i} {\dright A\over \partial \pi^i}  \label{FPB2}
\en
where $[\phi_i] [\pi^i]$ is the scalar product:
\eq
[\phi_i] [\pi^i] \equiv p_i (d-p_i -1) + \eta_i \eta_i = p_i d + \eta_i ~ (mod~2)
\en
Similarly, formula (\ref{leftright}) can be written as
\eq
{\dleft F \over \partial A} = (-)^{[a]  [f+1]} ~{\dright F \over \partial A} \label{leftright2}
\en
with $[a] = (a, \eta_a)$ and $[f+1] = (f+1, \eta_f +1)$.

\subsection{Properties of the form Poisson bracket}

Using the definition (\ref{FPB}), the following relations for the FPB of  (\ref{FPB2}) can be shown to hold:
\eqa
& & \{ B,A \} = - (-)^{(a+d+1)(b+d+1)+ \eta_a \eta_b} \{ A,B \}  \label{prop1} \\
& & \{A,BC \} = B \{A,C \} + (-)^{c(a+d+1)+\eta_c \eta_a} \{A,B \} C  \label{prop2}\\
& & \{AB,C \} =  \{A,C \} B + (-)^{a(c+d+1)+ \eta_a \eta_c}  A \{B,C \}  \label{prop3} \\
& & (-)^{(a+d+1)(c+d+1)+\eta_a \eta_c} \{ A, \{ B,C \} \} + cyclic~=0\\
& & (-)^{(a+d+1)(b+d+1)+\eta_a \eta_b} \{  \{ B,C \},A \} + cyclic~=0 \label{prop5}
\ena
i.e. graded antisymmetry, derivation property, and form-Jacobi identities.  Here too the signs can be expressed
in vector notation, for example:
\eq
(-)^{(a+d+1)(b+d+1)+ \eta_a \eta_b} = (-)^{[a+ \phi_i + \pi^i ] [ b+ \phi_i + \pi^i]}
\en
since
\eq
[\phi_i + \pi^i] = (p_i + d - p_i -1, \eta_i + \eta_i) = (d+1,0)~~~(mod~2)
\en

\subsection{Infinitesimal canonical transformations}

We can define the action of infinitesimal form-canonical transformations on any $a$-form $A$ as follows:
\eq
\delta A =  \{A,G \}
\en
where $G=G(\phi_i, \pi^i)$ is a bosonic $(d-1)$-form, the generator of the canonical transformations. Then $ \{A,G \}$ is a $a$-form like $A$. The generator $G$ can contain bosonic or fermionic infinitesimal parameters $\epsi (x)$ depending only on spacetime, for example as $G = \epsi G'$. Then if the parameter $\epsi$ is bosonic (fermionic), $G'$ is bosonic (fermionic), so that $G=\epsi G'$ is always bosonic. 
These transformations preserve the canonical FPB relations (\ref{canonicalPB}), and therefore we can call them form-canonical transformations. As in the usual case the proof involves the Jacobi identities applied to $\phi_i, \pi^j, G$:
\eq
\{ \{ \phi_i, \pi^j \}, G \} + (-)^{p_i (p_i + d+1)+ \eta_i \eta_j} ~\{ \{ \pi^j , G\}, \phi_i \} + \{  \{ G, \phi_i \}, \pi^j  \} =0
\en
Using the graded antisymmetry of the FPB this reduces to:
\eq
 \{\phi_i ,  \{ \pi^j , G \} \} +  \{ \{\phi_i ,  G \} , \pi^j \}  =  \{ \{ \phi_i, \pi^j \},  G \} =  0
\en
since $\{ \phi_i, \pi^j\}  = \delta_i^j$ is a number. Then
\eqa
& & \{\phi'_i, \pi'^j \} = \{\phi_i +  \{ \phi_i, G \}, \pi^j +  \{ \pi^j ,  G \} \}  \nonumber \\
& & ~~~~~~~~~~~ = \{\phi_i , \pi^j  \}  +   \{\phi_i ,  \{ \pi^j ,  G \} \} +   \{ \{\phi_i , G \} , \pi^j \}  + O(\epsi^2) \nonumber \\
& & ~~~~~~~~~~~ = \{\phi_i , \pi^j  \}  + O(\epsi^2)
\ena
Q.E.D.

\subsection{Form-canonical algebras}

As discussed in \cite{CD}, the commutator of two infinitesimal canonical transformations generated by the ($d-1$)-forms $G_1$ and $G_2$ is again 
an infinitesimal canonical transformation, generated by the ($d-1$)-form  $\{ G_1,G_2 \}$. This is due to
\eq
\{ G_1, G_2 \} = -\{  G_2, G_1 \} 
\en
for ($d-1$)-form entries, and to the form-Jacobi identity
\eq
\{ \{ A,G_1 \}, G_2 \} - \{ \{ A,G_2 \}, G_1 \} = \{ A, \{ G_1, G_2 \} \}
\en
holding for any $p$-form $A$. Therefore the form-canonical transformations close an algebra.
This algebra is finite dimensional if all fundamental fields (``positions and momenta") are $p$-forms
with $p \ge 1$, since there is only a finite number of $(d-1)$-form polynomials made out of the fundamental fields.
On the other hand, if there are fundamental $0$-forms, the algebra becomes infinite dimensional because there
are infinitely many $(d-1)$-form polynomials. 

\subsection{Action invariance and Noether theorem} 

Consider the action 
\eq
S=\int_{\Mcal^d} d \phi_i~\pi^i - H
\en
Its variation under an infinitesimal form-canonical transformation generated by a ($d-1$)-form $G$ is
\eq
\delta S = \int_{\Mcal^d} d (\{ \phi_i , G \} ) \pi^i + d \phi_i \{ \pi^i , G \} - \{ H, G \} 
\en
and with a little algebra (see \cite{CD}) one finds:
\eq
\delta S =  \int_{\partial \Mcal^d}  (\{ \phi_i , G \}  \pi^i  -  G  ) -  \int_{\Mcal^d} \{ H, G \}  \label{Noether1}
\en
after using the definition of FPBs and 
\eq
dG =  d \pi^i~ {\dright G \over \partial \pi^i} + d \phi_i ~ {\dright G \over \partial \phi_i} 
\en
Thus the action is invariant (up to a boundary term) under the infinitesimal canonical form-transformation generated by $G$ iff 
\eq
\{ H, G \} =0
\en
up to a total derivative. This result reproduces Noether's theorem in form language, and is identical
to the result obtained in \cite{CD} for the purely bosonic case (fermionic signs cancel out).
\sk
Suppose now that the $d-1$-form generators include infinitesimal parameters and their exterior derivatives as follows:
\eq
\Gbb= \epsi (x) G + (d \epsi) F 
 \en
 where $G$ and $F$ are respectively a $g$-form and a $(g-1)$-form,  while 
  the parameters $\epsi (x)$ are $(d-g-1)$-form external fields depending only on spacetime.
 Then under the infinitesimal canonical transformation generated by $\Gbb$ the action varies as:
\eqa
& & \delta S =   \int_{\partial \Mcal^d}  \epsi (\{ \phi_i , G \}  \pi^i  - (-)^{g+d+1}  G  ) + d\epsi (\{ \phi_i , F \}  \pi^i  - (-)^{g+d} F ) \nonumber \\
& & ~~~~~~ +  \int_{\Mcal^d} d \epsi~ ((-)^{g+d+1} G - \{H,F \}) -  \epsi \{ H, G \}  \label{Noether3}
\ena
Thus $\epsi (x) G + (d \epsi) F $ is a gauge generator leaving the action invariant iff
\eq
(-)^{g+d+1}G -\{H,F \} =0,~~~\{H,G \} =0 \label{conditions2}
\en
{\bf Note 1:} invariance of the action under local symmetries implies hamiltonian constraints, as can be seen from the conditions (\ref{conditions2}). In fact part of these constraints are the primary constraints
due to momenta definitions. Equalities valid modulo constraints are said to be {\it weak} equalities, and are
wtitten with a $\approx$ symbol. Thus both equalities in (\ref{conditions2}) should be written with $\approx$.
\sk
\noi {\bf Note 2:}  the infinitesimal transformation generated by $\epsi (x) G + (d \epsi) F $ must preserve the constraints, implying
\eq
\{ constraints, G \} \approx 0,~~~\{ constraints, F \} \approx 0  \label{conditions2bis}
\en
{\bf Note 3:} as observed in \cite{CD}, the conditions (\ref{conditions2}) and (\ref{conditions2bis}) generalize to geometric theories with fundamental bosonic and fermionic $p$-form fields the conditions for gauge generators found in \cite{SCHS}, and provide the basis for a constructive algorithm yielding all the gauge generators. This procedure is applied in the next Sections.
\sk
\noi {\bf Note 4 :} $F$ and $G$ must be first-class quantities, i.e. have weakly vanishing FPS's with all the constraints,
but do not have necessarily to be constraints.


\sect{Anti de Sitter $N=1$ supergravity in $d=3$}


\subsection{Lagrangian and symmetries}

We consider here $N=1$ AdS supergravity in $d=3$ (see \cite{MS,GGRS,RuizPvN}, ref. \cite{AT} for its derivation as a super Chern-Simons theory, and ref. \cite{if3} for its group manifold construction).
The fields $\phi_i$  are the $d=3$ vierbein $V^a$, the spin connection $\om^{ab}$ and the Majorana gravitino $\psi$. The 3-form Lagrangian is
\eq
L (\phi, d\phi) = R^{ab} V^c \epsi_{abc}  + 2 i \psibar \Sigma + {2 \over 3 \lambda^2} V^a V^b V^c - {i \over 2 \lambda} \psibar \gamma^{ab} \psi V^c \epsi_{abc} 
\label{EHLagrangiand3}
\en
where the super $AdS$ curvatures are defined as 
\eqa
& & R^a = dV^a -\om^a_{~b} ~ V^b - {i \over 2} \psibar \gamma^a \psi  \\
& & R^{ab}= d \om^{ab} - \om^a_{~e}~ \om^{eb} -{1 \over \lambda^2} V^a V^b + {i \over 2 \lambda} \psibar \gamma^{ab} \psi \\
& & \Sigma = d \psi -{1 \over 4} \omega^{ab} \gamma_{ab} \psi - {1 \over 2 \lambda} V^a \gamma_a \psi 
\ena
and $\lambda$ is the AdS radius. The supertorsion $R^a$ does not appear in the Lagrangian. The field equations are simply $R^a=R^{ab}=\Sigma =0$.

The action is invariant under the $d=3$ anti-De Sitter supergroup $OSp(1|2) \times Sp(2)$,
with the following gauge transformations on the basic fields:
\eqa
& & \delta V^a = \Dcal \epsi^a + \epsi^a_{~b} V^b + i \epsilonbar \gamma^a \psi \label{gauged3V}\\
& & \delta \omega^{ab} = \Dcal \epsi^{ab} - {2 \over \lambda^2} V^{[a} \epsi^{b]} - {i \over \lambda} \epsilonbar \gamma^{ab} \psi \\
& & \delta \psi = \Dcal \epsilon - {1 \over 2 \lambda} V^a \gamma_a \epsilon + {1 \over 4} \epsilon^{ab} \gamma_{ab} \psi + {1 \over 2\lambda} \epsi^a \gamma_a \psi \label{gauged3psi}
\ena
where the gauge parameters $\epsi^a$, $\epsi^{ab}$ and $\epsilon$ correspond to translations, Lorentz rotations and supersymmetry
transformations, respectively. These symmetry variations will be recovered in the  form-canonical treatment that follows.  

The action corresponding to (\ref{EHLagrangiand3})  is also invariant under the transformations:
  \eqa
         & & \delta V^a = \delta_{gauge} ~V^a  \label{LieVa2}\\
& & \delta \omega^{ab} = \delta_{gauge}~ \omega^{ab}  + 2 \epsi^c R^{ab}_{~~cd} V^d + \thetabar^{ab}_{~~c} \epsilon V^c
  - \thetabar^{ab}_{~~c}  \psi \epsi^c\\
& & \delta \psi = \delta_{gauge} ~\psi + 2 \epsi^a \Sigma_{ab} V^b \label{Liepsi2}
\ena 
where $\delta_{gauge}$ refers to the variations in (\ref{gauged3V})-(\ref{gauged3psi}), the components $R^{ab}_{~~cd}$ and
$ \Sigma_{ab}$ are defined as 
\eq
R^{ab} = R^{ab}_{~~cd} ~V^c V^d ,~~~\Sigma = \Sigma_{ab} V^a V^b 
\en
and
 \eq
         \thetabar^{ab}_{~~c} = 2i ~\Sigmabar_c^{~[a} \gamma^{b]}- i~ \Sigmabar^{ab} \gamma_c
         \en
         These transformations can be interpreted as superdiffeomorphisms on the super AdS manifold. They close only on-shell,
         and in order to promote them to symmetries that close also off-shell one has to introduce auxiliary fields, see for ex. 
         the review \cite{LC2018}, and ref.s \cite{if3,CCG}.

\subsection{Form hamiltonian and constraints}

The 1-form momenta conjugated to $V^a$, $\omega_{ab}$ and $\psibar$ are respectively :
\eqa
& & \pi_{a} = {\partial L \over \partial (dV^a)} = 0 \\
& & \pi_{ab} = {\partial L \over \partial (d \om^{ab})} =V^c  \epsi_{abc} \\
& & \pi =  {\partial L \over \partial (d \psibar)} =  2 i \psi
\ena
All momenta definitions are {\it primary constraints}:
\eq
\Phi_a \equiv \pi_a = 0,~~~\Phi_{ab} \equiv \pi_{ab} - V^c \epsi_{abc} = 0,~~~\Phi \equiv \pi - 2i\psi=0
\en
 since they do not involve the ``velocities" $dV^a$, $d\om^{ab}$ and $d \psi$.  The 3-form Hamiltonian is:
\eqa
& & H= dV^a ~ \pi_a + d \om^{ab}~ \pi_{ab} + d \psibar \pi - L = \\
& & ~~~ = dV^a ~ \Phi_a + d \om^{ab}~ \Phi_{ab}  + d \psibar ~\Phi + \om^a_{~d} ~\om^{db}~ V^c  \epsi_{abc}  + {i \over 2} \psibar ~\omega^{ab} \gamma_{ab} \psi \nonumber \\
& & ~~~~~~~ + {i \over \lambda} \psibar ~V^a \gamma_a \psi + {1 \over 3 \lambda^2} V^a V^b V^c 
\ena
The Hamilton equations of motion for $dV^a$, $d\omega^{ab}$ and $d\psi$
are identities,  while for the momenta they read:
\eqa
& & d \pi_a =  {\partial H \over \partial V^a} = -  R^{bc}  \epsilon_{abc}  \label{dpia}\\
& & d\pi_{ab} =  {\partial H \over \partial \omega^{ab}} = 2 \omega^c_{~[a} V^d  \epsilon_{b]cd} - {i \over 2} \psibar \gamma^{ab} \psi \\
& & d\pi = - 4 i ~\Sigma + 2i~ d\psi \label{dpi}
\ena
Requiring the ``conservation" of
$\Phi_a$ and $\Phi_{ab}$ leads to the
conditions:
\eqa
& & d \Phi_a = \{ \Phi_a,H \} = - R^{bc} \epsi_{abc}  = 0  \label{secondary1d3} \\
& & d \Phi_{ab} = \{ \Phi_{ab},H \} = - R^c \epsi_{abc}  = 0 \label{secondary2d3} \\
& & d \Phi = \{ \Phi , H \} = - 4i~\Sigma  = 0 \label{secondary3d3} 
\ena
implying the vanishing of all curvatures: $R^a =0$, $R^{ab}=0$, $\Sigma = 0$.  These are the equations of motion of $d=3$ 
anti-De Sitter supergravity, and completely determine the ``velocities" $dV^a$, $d\omega^{ab}$ and $d \psi$:
\eq
dV^a=\omega^a_{~b} ~V^b,~~~d \omega^a_{~b} = \omega^a_{~c}~ \omega^{cb},~~~d \psi = {1 \over 4} \omega^{ab} \gamma_{ab} \psi + {1 \over 2 \lambda} V^a \gamma_a \psi \label{velocitiesfixed}
\en
 Using the form Poisson bracket we find the constraint algebra:
  \eq
 \{ \Phi_a,\Phi_{bc} \}=-\epsi_{abc},~~~\{ \Phi^\alpha,\Phi^\beta \} = 4i~C^{\alpha\beta}
  \en
\noi  all other FPB's vanishing;  $C^{\alpha\beta}$ is  the charge conjugation matrix. 
Thus constraints are second-class, and this is consistent with the fact that all the  ``velocities" 
get fixed by requiring conservation of the primary constraints. The three constraints $\Phi_{ab}$ ($ab=12,13,23$) are equivalent
to the three linear combinations $\Xi^a= {1 \over 2} \epsilon^{abc} \Phi_{bc}$, and we find
\eq
\{ \Phi_a, \Xi^b \} = \delta^b_a
\en
We'll use the $\Xi^a$ in the definition of Dirac brackets of next Section. Note that form-Poisson brackets between bosonic (fermionic) 1-forms are symmetric (antisymmetric) in $d=3$, and in all odd dimensions, see eq. (\ref{prop1}). Also, the FPB betwen constraints yield numbers in $d=3$ gravity, and this allows a definition of form-Dirac brackets (see next Section). A similar definition is not available in $d=4$, since the FPB between constraints yield 1-forms, and the corresponding FPB matrix has no obvious inverse. 
\sk
\noi {\bf Note:} the action variations  (\ref{Noether1}) and (\ref{Noether3}) have been deduced assuming that $H$ depends only on
basic fields and momenta. This is not the case in constrained systems, where some of the velocities remain undetermined,
and therefore appear in the hamiltonian. However they always appear multiplied by primary constraints, and the
variation of these terms always vanishes weakly.

\subsection{Form Dirac brackets}

We define form Dirac brackets as follows
\eq
\{f,g \}^* \equiv \{f,g \} - \{f, \Phi_a \} \{ \Xi^a, g \} - \{f, \Xi^a \} \{\Phi_a,g \}- {1 \over 4i} \{ f, \Phibar_\alpha \} \{ \Phi^\alpha,g \}
\en
These Dirac brackets vanish strongly if any entry is a constraint $\Phi_a$, $\Xi^a$ or $\Phi^\alpha$. With the help of the general formulas
(\ref{prop1})-(\ref{prop5}) with $d=3$ we can verify that the Dirac brackets inherit the same properties of the Poisson brackets,
i.e. :
\eqa
& & \{ B,A \}^* = - (-)^{ab + \eta_a \eta_b} \{ A,B \}^*  \label{prop1d3} \\
& & \{A,BC \}^* = B \{A,C \}^* + (-)^{ca+\eta_c \eta_a} \{A,B \}^*C \label{prop2d3}\\
& & \{AB,C \}^* =  \{A,C \}^* B + (-)^{ac + \eta_a \eta_c}  A \{B,C \}^* \label{prop3d3} \\
& & (-)^{ac + \eta_a \eta_c} \{ A, \{ B,C \}^* \}^* + cyclic~=0\\
& & (-)^{ab+ \eta_a \eta_b} \{  \{ B,C \}^*,A \}^* + cyclic~=0
\ena
Using Dirac brackets the second-class constraints (i.e. all the constraints of the $d=3$ theory) disappear from the game, and 
we can use the 3-form Hamiltonian
\eq
H =   \om^a_{~e} ~\om^{eb}~ V^c  \epsi_{abc} + {i \over 2} \psibar \omega^{ab} \gamma_{ab} \psi + {i \over \lambda} \psibar V^a \gamma_a \psi + {1 \over 3 \lambda^2} V^a V^b V^c \epsi_{abc}
\en
The nonvanishing Dirac brackets between the basic fields and their momenta are given by:
\eqa
& & \{ V^a, \omega^{bc} \}^* = - {1 \over 2} \epsilon^{abc},~~~\{ \omega^{ab}, \pi_{cd} \}^*= \delta^{ab}_{cd},~~~\{ \psibar_\alpha, \pi^\beta \}^*= {1 \over 2} \delta^\beta_\alpha \\
& &   \{ \psibar_\alpha, \psi^\beta \}^* = {1 \over 4i} \delta^\beta_\alpha,~~~ \{ \psi^\alpha, \psi^\beta \}^* = -{1 \over 4i} \delta^\beta_\alpha,~~~ \{ \pi^\alpha, \pi^\beta \}^* = -i C^{\alpha\beta}
\ena

The Hamilton equations expressed via the Dirac bracket become:
\eqa
& & dV^a = \{ V^a, H \}^* = \omega^a_{~b} V^b + {i \over 2} \psibar \gamma^a \psi~~\Rightarrow R^a=0 \\
& & d \omega^{ab}= \{ \omega^{ab}, H \}^*  = \omega_e^{~[a} \omega^{b]e} - {i \over 2\lambda} \psibar \gamma^{ab} \psi + {1 \over  \lambda^2}  V^a V^b~~\Rightarrow R^{ab}=0 \nonumber\\
& & d \psi = \{ \psi, H \}^* = { 1 \over 4} \omega^{ab} \gamma_{ab} \psi + {1 \over 2 \lambda} V^a \gamma_a \psi   ~~ \Rightarrow \Sigma=0
\ena
i.e. the field equations of $d=3$ AdS supergravity. For the ``evolution" of the momenta we find:
\eqa
& & d \pi_a = \{ \pi_a, H \}^* = 0 ~~~\Leftrightarrow ~~d \Phi_{a}=0 \label{dpia2} \\
& & d \pi_{ab} =  \{ \pi_{ab}, H \}^* =  \epsilon_{abc} \omega^c_{~d} V^d - {i \over 2} \psibar \gamma_{ab} \psi~~~\Leftrightarrow ~~d \Phi_{ab}=0\\
& & d \pi= \{ \pi, H \}^* = {i \over 2} \omega^{ab} \gamma_{ab} \psi + {i \over \lambda} V^a \gamma_a \psi~~~~~~\Leftrightarrow ~~d \Phi =0 \label{dpi2}
\ena
where in the second line we used the identity 
\eq
\omega_{[a}^{~d} \epsilon_{bc]d} =0
\en
The momenta evolutions re-express the fact that the constraints are conserved, or equivalently that the exterior derivative of the momenta
is in agreement with their expression given by the second-class constraints.
\sk
\noi {\bf Note:} as observed after eq. (\ref{velocitiesfixed}), the field equations fix all the undetermined velocities $dV^a$, $d\omega^{ab}$ and $d \psi$. On the other hand, using Dirac brackets really amounts to substitute these ``velocities" with their expressions in terms of canonical variables given in (\ref{velocitiesfixed}), so that eq.s 
(\ref{dpia2})-(\ref{dpi2}) can be obtained from (\ref{dpia})-(\ref{dpi}) by setting $R^{ab}=0$, $\Sigma =0$.

\subsection{Canonical gauge generators}

Now we apply our procedure to find the canonical generators for Lorentz transformations, gauge translations and supersymmetry. 
\sk
\subsubsection{Lorentz gauge rotations}
\sk
We start from the first class 1-forms $\pi_{ab}$. They are first class in the since they have vanishing Dirac brackets
with all the constraints. Actually the constraints being all second class, they have been effectively eliminated from the theory 
by the use of Dirac brackets.  We take these 1-forms $\pi_{ab}$ as the ($d-2$)-forms $F$  in eq. (\ref{conditions2}), and find the ($d-1$)-forms $G$ that complete the gauge generator:
\eq
G_{ab} = \{H,F_{ab} \}^* = \{ H, \pi_{ab}\}^*  =  2 \omega^c_{~[a} V^d  \epsilon_{b]cd}  - {i \over 2} \psibar \gamma_{ab} \psi \label{condition3}
\en 
Next we have to check that $\{H,G_{ab} \}=0$. Notice that here it is useless to add to $G_{ab}$ any combination of constraints, since 
second-class constraints have no effect in a generator when using Dirac brackets. So $\{H,G_{ab} \}^*=0$ must hold with the $G_{ab}$ as given in (\ref{condition3}), and indeed this is the case, as one can check with a little algebra.

Thus
\eq
\Gbb = d\epsi^{ab} F_{ab} + \epsi^{ab} G_{ab} =  d\epsi^{ab} \pi_{ab} + \epsi^{ab}  (2 \omega^c_{~[a} V^d  \epsilon_{b]cd}
- {i \over 2} \psibar \gamma_{ab} \psi)
\en
generates gauge transformations via the Dirac bracket. Using the (second-class) constraint $\pi_{ab}= \epsilon_{abc} V^c$ in the
second term of the generator yields
\eq
\Gbb=d\epsi^{ab} \pi_{ab} + 2 \epsi^{ab}  \omega^c_{~[a} \pi_{b]c} - {i \over 2} \epsi^{ab} \psibar \gamma_{ab} \psi=  (\Dcal \epsi^{ab} ) \pi_{ab} - {i \over 2} \epsi^{ab} \psibar \gamma_{ab} \psi
\en
It generates local Lorentz transformations with parameter $\epsilon_{ab} (x)$, since
\eqa
& & \delta V^a = \{V^a, \Gbb \}^* = 2 \{\omega^{[b}_{~~d}, V^a \}^* \epsi^{c]d} \pi_{bc} = \epsi^a_{~b} V^b \\
& & \delta \omega^{ab} = \{ \omega^{ab}, \Gbb \}^* = \Dcal \epsi^{ab} \\
& & \delta \psi = \{ \psi, \Gbb \}^* =\{ \psi, -{i\over 2} \psibar \gamma_{ab} \psi \}^*  \epsi^{ab} =  {1 \over 4}  \epsi^{ab}\gamma_{ab} \psi \\
& & \delta \pi_a = \{ \pi_a, \Gbb \}^* =0 \\
& &  \delta \pi_{ab} = \{ \pi_{ab}, \Gbb \}^* =  \{ \epsilon_{abc} V^c, \Gbb \}^* =  \epsi_{~[a}^{c} \pi_{b] c} \\
& &  \delta \pi = \{ \pi, \Gbb \}^* =  \epsi^{ab} \{ \pi, -{i\over 2} \psibar \gamma_{ab} \psi \}^* =  {1 \over 4}  \epsi^{ab} \gamma_{ab} \pi
\ena
Note that  $\delta \pi_a = 0$ since $\Gbb$ has no effect on second class constraints.
\sk
\subsubsection{Gauge translations}
\sk
The procedure of the preceding paragraph can be started with any 1-form: indeed here any 1-form has
vanishing Dirac brackets with the constraints. We choose $F_a$ to be $\epsilon_{abc} \omega^{bc}$, since
this 1-form is conjugated to $V^a$, and therefore a good candidate to multiply the $d \epsi^a$ term in 
the generator of the gauge translations. Then $G_a$ is found in the usual way:
\eq
G_a =  \{H, F_a \}^* =  \epsilon_{abc} ( \omega^b_{~d} \omega^{dc}-{i \over 2\lambda} \psibar \gamma^{bc} \psi + {1 \over \lambda^2} V^b V^c)
\en
We have now to check that the second condition in (\ref{conditions2}) is satisfied, i.e. that $\{ H, G_a \}^* =0$. This condition gives
rise to four structures: $\omega \omega \omega$, $\omega \omega V$, $\psi \psi \omega$ and $\psi \psi V$. By explicit computation 
one can verify that the coefficients of these four terms are all zero (the $\omega\omega\omega$ term was shown to vanish in ref. \cite{CD}).

Therefore 
\eq
\Gbb = d \epsi^a F_a + \epsi^a G_a =  ({\cal D} \epsi^a) \epsi_{abc} \omega^{bc}+ \epsi^a \epsi_{abc} (-{i\over 2\lambda} \psibar \gamma^{bc} \psi + {1 \over \lambda^2} V^b V^c)
\label{diffgenerator}
\en
generates a symmetry. Its action on the basic fields is given by:
\eqa
& & \delta V^a = \{ V^a, \Gbb \}^* = \Dcal \epsi^a  \label{d3diff1}\\
& & \delta \omega^{ab}= \{ \omega^{ab} , \Gbb \}^* = {1 \over \lambda^2} (\epsi^a V^b - \epsi^b V^a) \\
& & \delta \psi= \{ \psi, \Gbb \}^*= {1 \over 2\lambda} \epsi^a \gamma_a \psi \\
& & \delta \pi_a = \{ \pi_a, \Gbb \}^*=0 \\
& & \delta \pi_{ab} = \{ \pi_{ab} , \Gbb \}^*  = \epsilon_{abc} \Dcal \epsi^c \\
& & \delta \pi= \{ \pi, \Gbb \}^*= {i \over \lambda} \epsi^a \gamma_a \psi={1 \over 2\lambda} \epsi^a \gamma_a \pi  \\ \label{d3diff4}
\ena
and reproduces the gauge translations contained in the variations (\ref{gauged3V})-(\ref{gauged3psi}).
This infinitesimal transformation clearly differs from the diffeomorphisms transformation in (\ref{LieVa2})-(\ref{Liepsi2}), even in second
order formalism (i.e. on the ``partial shell" $R^a=0$). Thus $\Gbb$ does not generate the {\it bona fide} infinitesimal diffeomorphisms one 
obtains acting with the Lie derivative.
\sk
\subsubsection{Gauge supersymmetry}
\sk
Another first-class 1-form is the fermionic momentum $\pi$. Taking $2\pi$ as candidate $F$ for a supersymmetry generator of the
form $\Gbb = d\epsilonbar F + \epsilonbar G$, we must have 
\eq
G =2 \{H,\pi\}^* = i \omega^{ab} \gamma_{ab} \psi + {2i \over \lambda} V^a \gamma_a \psi
\en
and check that $\{H,G \}=0$. This FPB yields two structures, $\omega\omega\psi$ and $\omega V \psi$, and both
have vanishing coefficient, as an explicit computation can verify. Therefore the canonical generator of gauge supersymmetry is:
\eq
\Gbb = 2d\epsilonbar~ \pi + i \epsilonbar~ \omega^{ab} \gamma_{ab} \psi + {2i\over\lambda} \epsilonbar~V^a \gamma_a \psi
=2 \Dcal \epsilonbar~ \pi+  {2i\over\lambda} \epsilonbar~V^a \gamma_a \psi
\en
where we used $\pi=2i\psi$ in the last equality. 

The gauge supersymmetry variations are
\eqa
& & \delta V^a = \{ V^a, \Gbb \}^* =  \{V^a, i \epsilonbar~\omega^{cd} \gamma_{cd} \psi \}^* = i \epsilonbar ~ \gamma^a \psi \\
& & \delta \omega^{ab} = \{ \omega^{ab}, \Gbb \}^* = 2 \{\omega^{ab}, {i \over \lambda}  \epsilonbar~V^a \gamma_a \psi \}^* =- {i \over \lambda}  \epsilonbar ~ \gamma^{ab} \psi \\
& & \delta \psi  = \{ \psi, \Gbb \}^* =  2\{\psi, \Dcal \epsilonbar ~\pi  \}^* = \Dcal \epsilon - {1 \over  2\lambda}~ V^a \gamma_a \epsilon \\
& & \delta \pi_a = \{ \pi_a, \Gbb \}^* =  0 \\
& & \delta \pi_{ab} = \{ \pi_{ab}, \Gbb \}^* = \{ V^c \epsi_{abc} , \Gbb \}^* = -i \epsilonbar~\gamma_{ab} \psi \\
& & \delta \pi= \{ \pi, \Gbb \}^* = \{2i\psi,\Gbb \}^* = 2i \Dcal \epsilon - {i \over \lambda} V^a \gamma_a \epsilon
\ena
and coincide (on the fundamental fields) with those contained in (\ref{gauged3V})-(\ref{gauged3psi}). Here we have found also the
gauge supersymmetry variations of the momenta.


\sect{$N=1$ supergravity in $d=4$, new minimal model}


\subsection{Lagrangian}

 The theory was first constructed in ref. \cite{SW}, and recast
in the group manifold formalism in ref. \cite{DFTvN}.

The basic fields are the vierbein $V^a$, the spin connection $\omega^{ab}$, the (Majorana) gravitino $\psi$, and the auxiliary fields 
$A$ (1-form) and $T$ (2-form). The 12 bosonic off-shell degrees of freedom (6 for $V^a$, 3 for $A$ and 3 for $T$) are balanced
by the 12 fermionic degrees of freedom for the Majorana gravitino $\psi$. The curvatures are defined as:
 \eqa
   & & R^{ab}=d \omega^{ab} - \omega^a_{~c} ~ \omega^{cb} \\
   & & R^a=dV^a - \omega^a_{~b} ~ V^b -{i \over 2} \psibar \gamma^a \psi \equiv \Dcal V^a -{i \over 2} \psibar \gamma^a \psi\\
   & & \rho = d\psi - {1 \over 4} \omega^{ab} \gamma_{ab} \psi  - {i \over 2} \ga_5 \psi A \equiv \Dcal \psi 
   - {i \over 2} \ga_5 \psi A\\
   & & R^{\square} = dA \\
   & & R^{\otimes}=dT-{i \over 2} \psibar \gamma_a \psi ~V^a
\ena
  \noi Taking exterior derivatives of both sides yields the Bianchi identities:
     \eqa
    & &  \Dcal R^{ab} =0 \\
    & &  \Dcal R^a + R^a_{~b} ~ V^b - i~ \psibar \gamma^a \rho =0\\
    & & \Dcal \rho + {1 \over 2} \gamma_5 \rho A+ {1 \over 4} R^{ab} \gamma_{ab} ~\psi - {i \over 2} \gamma_5 \psi R^\square=0\\
    & & dR^\square=0 \\
    & & dR^{\otimes} - i~ \psibar \gamma_a \rho V^a + {i \over 2} \psibar \gamma_a \psi~ R^a = 0
     \ena

 The Einstein-Hilbert action is
 \eq
           S = \int_{M^4} R^{ab} V^c V^d \epsilon_{abcd} + 4 \psibar \gamma_5 \gamma_a  \rho V^a - 4 R^\square T
           \label{spacetimeaction}
            \en
            
     \subsection{Field equations}
               
               Varying $\omega^{ab}$, $V^a$, $\psi$, $A$, and $T$  in the action (\ref{spacetimeaction}) leads to the   
                equations of motion:
                \eqa
                 & & 2 \epsilon_{abcd} R^c V^d =0 ~~\Rightarrow ~R^a=0 \\
                 & & 2 R^{bc} V^d \epsilon_{abcd}-4 \psibar \gamma_5 \gamma_a \rho = 0 \\
                 & & 8 \gamma_5 \gamma_a \rho V^a - 4 \gamma_5 \gamma_a \psi R^a  =0 \\
                 & & R^\otimes = 0 \\
                 & &  R^\square  =0\\
                 \ena
        
         \subsection{Symmetries}   
         
         The action is invariant under diffeomorphisms (with parameter $\epsi^a$), supersymmetry ($\epsilon$), Lorentz rotations ($\epsi^{ab}$),
         $U(1)$ gauged by $A$ (parameter $\eta$) and an abelian symmetry gauged by the two-form $T$ (1-form parameter $\mu$). The symmetry variations are:
         \eqa
        & &  \delta V^a = \Dcal \epsi^a + 2 R^{a}_{bc} \epsi^b V^c + \epsi^a_{~b} V^b - i \psibar \gamma^a \epsilon \label{symd4V}  \\
        & & \delta \omega^{ab} = \Dcal \epsi^{ab} + 2 R^{ab}_{~~cd} \epsi^c V^d - \thetabar^{ab}_{~~c} \psi \epsi^c + \thetabar^{ab}_{~~c} \epsilon    
        V^c + 3 i  \epsi^{abcd} ~\psibar \gamma_c \epsilon \\
        & & \delta \psi = \Dcal \epsilon + {i \over 2} A \epsilon + {3i \over 4} \gamma_5 \epsilon f_a V^a - {3i \over  2}\gamma_5 \gamma_{ab} \epsilon V^a f^b + {1 \over 4} \gamma^{ab} \epsi_{ab} \psi+  \\
        & &~~~~~~~+ 2 \rho_{ab} \epsi^a V^b - {3i \over 4} \gamma_5 \psi f_a \epsi^a + {3i \over 2} \gamma_5 \gamma_{ab} \psi \epsi^a f^b - {i \over 2 } \gamma_5 \psi \eta\\
        & & \delta T =  d \mu - i \psibar \gamma^a \epsilon + {i \over 2} \epsi^a \psibar \gamma_a \psi + 3 f^a \epsi^b V^c V^d \epsi_{abcd} \\
        & & \delta A = d \eta + 2 F_{ab} \epsi^a V^b - \psibar \chi_a \epsi^a + \epsilonbar \chi_a V^a + {9i\over 2}\epsilonbar \gamma_a \psi f^a \label{symd4A}
         \ena     
            with $f_a$ defined as $R^\otimes = f^a V^b V^c V^d \epsi_{abcd} $,  and
            \eqa
            & & \thetabar^{ab}_{~~c} \equiv 2 i \rhobar_c^{~[a} \gamma^{b]} - i \rhobar^{ab} \gamma_c \\
            & & \chi_a \equiv 2 (\gamma_5 \gamma^b \rho_{ab} + {i \over 8}\epsi_{abcd} \gamma^b \rho^{cd}) 
            \ena
            These symmetries close off-shell, thanks to the auxiliary fields.
 
\subsection{Form hamiltonian and constraints}

The 2-form momenta conjugate to $V^a$, $\omega_{ab}$, $\psi$ and $A$, and the 1-form momentum conjugate to the 2-form $T$ are respectively\footnote{unless stated otherwise, all partial
derivatives act from the left in the following.} :
\eqa
& & \pi_{a} = {\partial L \over \partial (dV^a)} = 0 \\
& & \pi_{ab} = {\partial L \over \partial (d \om^{ab})} =V^c V^d \epsi_{abcd} \\
& & \pi =  {\partial L \over \partial (d \psibar)} = 4 \gamma_5 \gamma_a \psi V^a \\
& & \pi (A) =  {\partial L \over \partial (dA)} = - 4 T \\
& & \pi(T) =  {\partial L \over \partial (dT)}=0
\ena
All momenta definitions are {\it primary constraints}:
\eqa
& & \Phi_a \equiv \pi_a ,~~~\Phi_{ab} \equiv \pi_{ab} - V^cV^d \epsi_{abcd} ,~~~\Phi \equiv \pi-4\gamma_5 \gamma_a \psi V^a \\
& & \Phi(A) \equiv \pi(A)+4T,~~~\Phi(T) \equiv \pi(T)
\ena
The form Hamiltonian is:
\eqa
& & H= dV^a ~ \pi_a + d \om^{ab}~ \pi_{ab} + d\psibar ~\pi + dA ~\pi(A) + dT~ \pi(T)  \nonumber \\
& & ~~~~ - d \om^{ab} ~V^c V^d \epsi_{abcd} + \om^a_{~e} ~\om^{eb}~ V^c  V^d \epsi_{abcd} + 4 d\psibar \gamma_5 \gamma_a \psi V^a 
 \nonumber \\
 & & ~~~ -2i \psibar \gamma_a \psi A V^a -i\psibar \gamma^d \omega^{bc} V^a \epsi_{abcd} + 4 (dA)T 
 \ena
 or, using the definition of the constraints:
 \eqa
& & H = dV^a ~ \Phi_a + d \om^{ab}~ \Phi_{ab}  + d\psibar~ \Phi + dA~ \Phi(A) + dT ~\Phi(T) \nonumber \\
& & ~~~ + \om^a_{~e} ~\om^{eb}~ V^c V^d  \epsi_{abcd}  -2i \psibar \gamma_a \psi A V^a -i\psibar \gamma^d \omega^{bc} V^a \epsi_{abcd} 
\ena
The Hamilton equations giving $dV^a$, $d\om^{ab}$, $d\psi$, $dA$ and $dT$  are identities, so these ``velocities" are undetermined at this stage. The Hamilton equations for the momenta read:
\eqa
& & d \pi_a =  \{ \pi_a,H \} = - 2 R^{bc} V^d \epsilon_{abcd} + 4 \psibar \gamma_5 \gamma_a \rho\\
& & d\pi_{ab} = \{ \pi_{ab},H \}= 2 \omega^c_{~[a} V^d V^e \epsilon_{b]cde} + i \psibar \gamma^c \psi V^d \epsi_{abcd}  \\
& & d\pi = \{ \pi,H \}=  - 4 \gamma_5 \gamma_a d \psi V^a - 4i \gamma_a \psi A V^a - 2i \gamma^d \psi \omega^{bc} V^a \epsi_{abcd}\\
& & d\pi(A)= \{ \pi(A),H \}= -2i \psibar \gamma_a \psi V^a \\
& & d\pi(T)= \{ \pi(T),H \}= -4 dA 
\ena

Requiring the ``conservation" of the constraints leads to the
conditions:
\eqa
& & 0= d \Phi_a = \{ \Phi_a,H \} = -2 R^{bc} ~V^d \epsi_{abcd} + 4 \psibar \gamma_5 \gamma_a \rho  \\
& & 0 = d \Phi_{ab} = \{ \Phi_{ab},H \} =-2  R^c ~V^d \epsi_{abcd} \\
& & 0 = d \Phi = \{ \Phi,H \} =-8\gamma_5 \gamma_a \rho V^a + 4 \gamma_5 \gamma_a \psi R^a\\ 
& & 0 = d \Phi(A) = \{ \Phi(A),H \} =4 R^\otimes \\
& & 0 = d \Phi(T)= \{ \Phi(T),H \} =-4dA
\ena
that reproduce the field equations of the theory. 
Note that these conditions fix the velocities $dV^a$, $dA$ and $dT$ in terms of the canonical variables:
\eq
dV^a = \omega^a_{~b} V^b,~~~dA=0,~~~dT={i\over 2} \psibar \gamma^a \psi V^a
\en
while $d\omega^{ab}$ and $d\psi$ are constrained to satisfy the Einstein and Rarita-Schwinger field equations.
 
Using the form Poisson bracket we find the constraint algebra:
  \eqa
& & \{ \Phi_a,\Phi_{bc} \}=-2\epsi_{abcd} V^d,~~\{ \Phi_a,\Phi \}=4 \gamma_5 \gamma_a \psi,\\
& & \{ \Phi^\alpha,\Phi^\beta \}=8 (\gamma_5 \gamma_a C^{-1})^{\alpha\beta} V^a,~~\{ \Phi(A),\Phi(T) \}=4
  \ena
 all other PFB between constraints vanishing. Thus the constraints are not all first-class. This is consistent with the fact that some of the undetermined ``velocities"  get fixed by requiring conservation of the primary constraints. Classical references on constrained hamiltonian systems are given in
\cite{Dirac,HRT,HT}.

\subsection{Dirac brackets}

We can eliminate the second-class constraints $\Phi(A), \Phi(T)$ by using the Dirac brackets defined as:
\eq
\{ A,B \}^* \equiv \{ A,B \} + {1 \over 4} \{ A,\Phi(A) \}\{ \Phi(T),B \}- {1 \over 4} \{ A,\Phi(T) \}\{ \Phi(A),B \}
\en
and satisfy the relations (\ref{prop1})-(\ref{prop5}) with $d=4$. With the above definition $\{ \Phi(A), anything \}^* = \{ \Phi(T), anything \}^*=0$.

\subsection{Canonical gauge generators}
\sk
\subsubsection{Lorentz gauge transformations}
\sk
We start from the first class 2-forms $\pi_{ab}$, having vanishing Dirac brackets with all the constraints, 
and take them as the ($d-2$)-forms $F$  in eq. (\ref{conditions2}).  To find the corresponding ($d-1$)-forms $G_{ab}$ that complete the
gauge generator one uses the first condition in (\ref{conditions2}), yielding $G_{ab}$ as the Dirac bracket of $H$ with $F_{ab}$, up to constraints. Since
\eq
\{H,\pi_{ab} \}^* = 2 \omega_{~[a}^{e} V^c V^d \epsilon_{b]ecd}+i \psibar \gamma^c \psi V^d \epsi_{abcd}
\en
we find that
\eq
G_{ab} = 2 \omega_{~[a}^{e} V^c V^d \epsilon_{b]ecd}+i \psibar \gamma^c \psi V^d \epsi_{abcd}+ \alpha_{ab}^c ~ \Phi_c + \beta_{ab}^{cd} ~\Phi_{cd} + \xibar_{ab} \Phi + \Phibar \zeta_{ab}
\en
\noi where $\alpha_{ab}^c $, $\beta_{ab}^{cd} $, $\xibar_{ab}$ and $\zeta_{ab}$ are 1-form coefficients to be determined by the second condition in
(\ref{conditions2}), i.e. weak vanishing of the Dirac bracket between $H$ and $G_{ab}$. This yields
\eq
\alpha_{ab}^c = \delta^c_{[a} V_{b]},~~~\beta_{ab}^{cd} = 2 \omega_{[a}^{~~c}~ \delta^d_{b]},~~~\xibar=-{1 \over 8} \psibar \gamma_{ab},~~~
\zeta_{ab}= {1 \over 8}  \gamma_{ab} \psi
\en
so that $G_{ab}$ becomes:
\eq
G_{ab} =  2 \omega^c_{~[a} \pi_{b] c} - V_{[a} \pi_{b]} - {1 \over 4} \psibar \gamma_{ab} \pi
\en
It is easy to check that this $G_{ab}$ has weakly vanishing Dirac brackets with the constraints $\Phi_a$, $\Phi_{ab}$, $\Phi$ and is
therefore a first-class 3-form. 

We have thus constructed the gauge generator
\eqa
& & \Gbb = \epsi^{ab} G_{ab} + d \epsi^{ab} F_{ab} = \epsi^{ab} (2 \omega^c_{~a} \pi_{bc} - V_{a} \pi_{b} - {1 \over 4} \psibar \gamma_{ab} \pi) + (d \epsi^{ab}) \pi_{ab}  \nonumber \\
& & ~~  =\Dcal \epsi^{ab} \pi_{ab} - \epsi^{ab} V_a \pi_b - {1 \over 4} \psibar \gamma_{ab} \pi
\ena
It generates the Lorentz gauge rotations on all canonical variables. Indeed
\eqa
& & \delta V^a = \{V^a, \Gbb \}^* = \epsi^a_{~b} V^b, ~~\delta \omega^{ab} = \{\omega^{ab}, \Gbb \}^* = \Dcal \epsi^{ab} , ~~\delta \psi = \{\psi, \Gbb \}^*= {1 \over 4 } \epsi^{ab} \gamma_{ab} \psi \nonumber\\
& &  \delta \pi_a = \{\pi_a, \Gbb \}^* = \epsi_a^{~b} \pi_b , ~~\delta  \pi_{ab} =  \{\pi_{ab}, \Gbb \}^*  = \epsi_{~[a}^{c} \pi_{b] c},~~\delta \pi =  
\{\pi, \Gbb \}^* =  {1 \over 4 } \epsi^{ab} \gamma_{ab} \pi \nonumber \\
\ena
and satisfies all the conditions to be a symmetry generator of the action.

\subsubsection{U(1) transformations}

 Another first-class 2-form is $\pi(A)$, or equivalently $T$, since using the Dirac brackets 
 we can use the second-class constraint $\pi(A) + 4 T =0$ as a strong equality.  Starting the procedure 
 for constructing the gauge generator yields 
 \eq
 G= \{H,\pi(A) \}^* + constraints
 \en
 where the part proportional to constraints is determined by requiring
 \eq
 \{G,H \}^* \approx 0
 \en
 Using 
 \eq
 \{H,\pi(A) \}^* = -2 \psibar \gamma_a \psi V^a
 \en
 one finds for $G$:
 \eq
 G=- 2i \psibar \gamma_a \psi V^a + {i \over 2} \psibar \gamma_5 \Phi= { i \over 2} \psibar \gamma_5 \pi
 \en
 and it is easy to verify that $G$ has vanishing Dirac bracket with all the constraints (i.e. with $\phi_a$, $\phi_{ab}$ and $\phi$, the constraints
 $\phi(A)$ and $\phi(T)$ having strongly vanishing Dirac brackets with anything) . Then the generator
 \eq
 \Gbb = d\eta ~ \pi(A) + \eta {i\over 2} \gamma_5 \pi
 \en
 satisfies all conditions to be a gauge generator. Its action is nontrivial only on the fields $\psi$ and $A$, and
 on the momenta $\pi$:
 \eqa
 & & \delta \psi = \{ \psi, \Gbb \} = - {i \over 2} \eta   \gamma_5 \psi \\
 & & \delta A =   \{ A, \Gbb \} = d \eta \\
 & & \delta \pi =  \{\pi, \Gbb \} =  - {i \over 2} \eta   \gamma_5 \pi
 \ena
 and reproduce on $\psi$ and $A$ the $U(1)$ transfornations contained in eq.s (\ref{symd4V})-(\ref{symd4A}). In addition we find here also
 the $U(1)$ transformations on the momenta conjugate to $\psi$. Not surprisingly, these momenta have the same 
 $U(1)$ charge as the gravitino.
 
 \subsubsection{Transformation with 1-form parameter}

Finally, there is the first-class 1-form $\pi(T)$ that one can use to find a symmetry generator with a 1-form parameter,
according to the discussion in Section 2.  The procedure here is immediate, since $\{H,\pi(T) \}^* =0$, and therefore
\eq
\Gbb = d\mu ~\pi(T)
\en
generates the transformation on the 2-form auxiliary field $T$:
\eq
\delta T = d\mu \{T, \Gbb \}^* = d\mu
\en 
contained in the symmetry variations given in  (\ref{symd4V})-(\ref{symd4A}).

\sect{Conclusions}
We have presented the covariant hamiltonian treatment of geometric theories containing bosonic and fermionic $p$-forms, and applied it
to supergravity in $d=3$ and $d=4$. Using form language, besides built-in Lorentz covariance (no time direction singled out) and invariance under diffeomorphisms, a further bonus is given by a
considerable simplification in the analysis: for contrast one can compare it with the canonical treatment of first order tetrad gravity in ref.
\cite{CVF}.  Our formulation allows also an algorithmic procedure to find all gauge symmetry generators of the theory, including
generators with $p$-form parameters, appearing in the gauge transformations of $(p+1)$-form gauge fields.

\section*{Acknowledgement}

This research has been partially supported by Universit\`a del Piemonte Orientale research funds.

\appendix

\sect{$\gamma$ matrices in $d=2+1$}

\eq
 \ga_0 =
\left(
\begin{array}{cc}
  i &  0    \\
  0&  -i
\end{array}
\right),~~~\ga_1=
\left(
\begin{array}{cc}
  0&   1  \\
  1&     0
\end{array}
\right)
,~~~\ga_2=
\left(
\begin{array}{cc}
  0&   -i \\
  i&     0
\end{array}
\right)
\en

\eqa
& & \eta_{ab} =(-1,1,1),~~~\{\ga_a,\ga_b\}=2 \eta_{ab},~~~[\ga_a,\ga_b]=2 \ga_{ab}= -2 \epsi_{abc} \ga^c, \\
& & \epsi_{012} = - \epsi^{012}=1, \\
& & \ga_a^\dagger = \ga_0 \ga_a \ga_0,~~\ga_a^T= - C \ga_a C^{-1}, ~~C^T = -C, ~~C^2 = \onebold
\ena

\subsection{Useful identities}

\eqa
 & &\ga_a\ga_b= \ga_{ab}+\eta_{ab}= - \epsi_{abc} \ga^c + \eta_{ab}\\
 & &\ga_{ab} \ga_c=\eta_{bc} \ga_a - \eta_{ac} \ga_b -\epsi_{abc}\\
 & &\ga_c \ga_{ab} = \eta_{ac} \ga_b - \eta_{bc} \ga_a -\epsi_{abc}\\
 & &\ga_a\ga_b\ga_c= \eta_{ab}\ga_c + \eta_{bc} \ga_a - \eta_{ac} \ga_b - \epsi_{abc}\\
 & &\ga^{ab} \ga_{cd} = - 4 \de^{[a}_{[c} \ga^{b]}_{~~d]} - 2 \de^{ab}_{cd}
  \ena
 \noi where $\de^{ab}_{cd}
 = \unmezzo (\de^a_c \de^b_d - \de^a_d \de^b_c)$, and index antisymmetrizations in square brackets have weight 1.

\subsection{Fierz identity for two Majorana one-forms}

\eq
\psi \psibar = {1 \over 2} (\psibar \ga^a \psi ) \ga_a
\en
As a consequence 
\eq
\ga_a \psi \psibar \ga^a \psi =0   \label{Fierz3d}
\en

\sect{$\gamma$ matrices in $d=3+1$}
 
\eqa
& & \eta_{ab} =(1,-1,-1,-1),~~~\{\ga_a,\ga_b\}=2 \eta_{ab},~~~[\ga_a,\ga_b]=2 \ga_{ab}, \\
& & \ga_5 \equiv  -i \ga_0\ga_1\ga_2\ga_3,~~~\ga_5 \ga_5 = 1,~~~\epsi_{0123} = - \epsi^{0123}=1, \\
& & \ga_a^\dagger = \ga_0 \ga_a \ga_0, ~~~\ga_5^\dagger = \ga_5 \\
& & \ga_a^T = - C \ga_a C^{-1},~~~\ga_5^T = C \ga_5 C^{-1}, ~~~C^2 =-1,~~~C^T =-C
\ena

\subsection{Useful identities}
\eqa
 & &\ga_a\ga_b= \ga_{ab}+\eta_{ab}\\
 & & \ga_{ab} \ga_5 = - {i \over 2} \epsilon_{abcd} \ga^{cd}\\
 & &\ga_{ab} \ga_c=\eta_{bc} \ga_a - \eta_{ac} \ga_b +i \epsi_{abcd}\ga_5 \ga^d\\
 & &\ga_c \ga_{ab} = \eta_{ac} \ga_b - \eta_{bc} \ga_a +i \epsi_{abcd}\ga_5 \ga^d\\
 & &\ga_a\ga_b\ga_c= \eta_{ab}\ga_c + \eta_{bc} \ga_a - \eta_{ac} \ga_b +i \epsi_{abcd}\ga_5 \ga^d\\
 & &\ga^{ab} \ga_{cd} = i \epsi^{ab}_{~~cd}\ga_5 - 4 \de^{[a}_{[c} \ga^{b]}_{~~d]} - 2 \de^{ab}_{cd}
 \ena

 \subsection{Charge conjugation and Majorana condition}

\eqa
 & &   {\rm Dirac~ conjugate~~} \psibar \equiv \psi^\dagger
 \ga_0\\
 & &  {\rm Charge~ conjugate~spinor~~} \psi^c = C (\psibar)^T  \\
 & & {\rm Majorana~ spinor~~} \psi^c = \psi~~\Rightarrow \psibar =
 \psi^T C
 \ena

\subsection{Fierz identity for two spinor one-forms}
\eq
 \psi  \chibar = \unquarto [ (\chibar  \psi) 1 + (\chibar \ga_5  \psi) \ga_5 + (\chibar \ga^a  \psi) \ga_a + (\chibar \ga^a \ga_5  \psi) \ga_a \ga_5  - \unmezzo (\chibar \ga^{ab}  \psi) \ga_{ab}]
 \en
 \subsection{Fierz identity for two Majorana spinor one-forms}
 \eq
 \psi  \psibar = \unquarto [  (\psibar \ga^a  \psi) \ga_a  - \unmezzo (\psibar \ga^{ab}  \psi) \ga_{ab}]
 \en
 \noi As a consequence
 \eq
\ga_a \psi \psibar \ga^a \psi =0,~~~ \psi \psibar \ga^a \psi- \ga_b \psi \psibar \ga^{ab} \psi=0 \label{Fierz4d}
\en

\vfill\eject
\end{document}